\documentclass[aps,pra,showpacs,twocolumn,groupedaddress,amsmath]{revtex4}
\usepackage{graphics}
\usepackage{graphicx}
\newcommand \beq{\begin{eqnarray}}
\newcommand \eeq{\end{eqnarray}}
\newcommand \bea{\begin{eqnarray}}
\newcommand \eea{\end{eqnarray}}
\newcommand{\ve}[1]{\mbox{\boldmath $#1$}}
\def\simge{\mathrel{%
         \rlap{\raise 0.511ex \hbox{$>$}}{\lower 0.511ex \hbox{$\sim$}}}}
\def\simle{\mathrel{
         \rlap{\raise 0.511ex \hbox{$<$}}{\lower 0.511ex \hbox{$\sim$}}}}

\begin{document}
  \title{Non-uniform vortex lattices in inhomogeneous rotating Bose-Einstein
condensates}
\author{Gordon Baym,$^{a,b}$  C. J. Pethick,$^{a}$ S. Andrew Gifford,$^{b}$
and Gentaro Watanabe$^{a,c}$}
\affiliation{
$^{a}$NORDITA, Blegdamsvej 17, DK-2100 Copenhagen \O, Denmark
\\
$^{b}$Department of Physics, University of Illinois, 1110 W. Green St.,
Urbana, IL 61801
\\
$^{c}$The Institute of Chemical and Physical Research (RIKEN),
2-1 Hirosawa, Wako, Saitama 351-0198, Japan
}

\date{\today}

\begin{abstract}

    We derive a general framework, in terms of elastic theory, for describing
the distortion of the vortex lattice in a rotating Bose-Einstein condensate at
arbitrary rotation speed and determining the dependence of the distortion on
the density inhomogeneity of the system.  In the rapidly rotating limit, we
derive the energetics in terms of Landau levels, including excitation to
higher levels; the distortion depends on the excitation of higher levels as
well as on the density gradient.  As we show, the dominant effect of higher
Landau levels in a distorted lattice in equilibrium is simply to renormalize
the frequency entering the lowest Landau level condensate wave function --
from the transverse trap frequency, $\omega$, to the rotational frequency,
$\Omega$, of the system.  Finally, we show how the equilibrium lattice
distortion emerges from elastohydrodynamic theory for inhomogeneous systems.

\end{abstract}


\maketitle

\section{Introduction}

    Studies of vortices in rapidly rotating Bose-Einstein condensates
\cite{rotexp} have enabled one to look in detail at the structure of the
vortex lattice.  While, to a first approximation the vortices form a uniform
triangular lattice in the limit of a large number of vortices, $N_v$,
Refs.~\cite{anglin} and \cite {sheehy} have shown, for slow rotation, that
the vortex lattice undergoes a slight distortion, becoming more widely
spaced near the edge of the cloud.  The distortion observed later by
Coddington et al.  \cite{coddington} is in good agreement with theoretical
predictions.  The distortion expected in the fast rotation -- lowest Landau
level (LLL) -- limit in a harmonic trap was discussed analytically in Ref.
\cite{wbp} and numerically in Ref.~\cite{read}.  Our aim in this paper is to
produce a general framework for calculating vortex lattice distortions at
arbitrary rotation rates in terms of the energetics of the distorted lattice.

    The distortion of the vortex lattice is driven by the particle-density
gradient in the system.  Since the energy of a vortex increases with
increasing local density, vortices tend to be forced towards regions of lower
density.  For clarity, we focus on the directions transverse to the rotation
axis (${\hat {\bf z}}$), and assume the system to be uniform along the
rotation axis; the ready generalization to include structure in the direction
of the rotation axis does not change the results conceptually.  In Sec.\ II we
give a general discussion of the energy of a non-uniform vortex lattice.  This
is applied to the case of slow rotation in Sec.\ III and to the case of
rotation rates comparable to the trap frequency in Sec.\ IV.  Finally, in
Sec.\ V we explore the relationship between the present work and
elastohydrodynamic theory \cite{chandler,tkmodes,qhmodes,gifford} which
describes the long wavelength modes of the vortex lattice.  We take $\hbar=1$
throughout.

\vspace{36pt}

\section{Energetics of vortex displacements}

    We assume that the vortex displacements from a uniform lattice vary slowly
in space and are described by a smooth displacement field, ${\ve
\epsilon}\,({\bf r})$, and that the lattice rotates at angular frequency
$\Omega$.  We work primarily in the rotating frame, where the total energy,
$E' = E - \Omega L_z$, is generally a functional of the independent degrees of
freedom -- the displacements of the vortices, the smoothed density, $n({\bf
r})$, and the smoothed superfluid velocity in the rotating frame, which we
denote by ${\bf v}_R$.  [We denote the local density, with all its wiggles
near the vortices, by ${\hat n}({\bf r})$, and the microscopic superfluid
velocity by $\hat {\bf v}$.] In the final section we consider the energy as a
functional of all these variables.  However, in equilibrium, the superfluid
velocity is determined by the displacement of the vortices by the equation for
quantization of circulation.  In this case it is most convenient to regard the
energy $E'$, for a slow variation of the lattice distortion, as a functional
only of the smoothed density, $n({\bf r})$, and the smoothed vortex density,
$n_v({\bf r})$.  The important terms, as we shall see, are the first order
change in the energy with $\ve\epsilon$, which is present only for a spatially
non-uniform system, and the change in the kinetic energy of rotation induced
by a lattice distortion, which leads to a term second order in $\ve \epsilon$.

    Since a uniform triangular lattice does not minimize $E'$, the first order
variation of $E'$ under a distortion of the vortex lattice is non-vanishing
and of the form
\beq
 \delta E' = \int d^2r\ \delta n_v({\bf r}) Q(n,n_v).
 \label{EQ}
 \eeq
(The energy $Q$ is the analog for a vortex of the quasiparticle energy
in the Landau theory of Fermi liquids.)  To first order in $\ve \epsilon$,
\beq
   \delta n_v({\bf r}) = -n_v^0 \ve\nabla \cdot \ve \epsilon(\bf r)\,,
  \label{deltanv}
\eeq
where $n_v^0=m\Omega/\pi$ is the vortex density of the uniform lattice.
Integrating by parts, we have
\beq
    \delta^{(1)} E' = \int n_v^0 \ve\epsilon \cdot \ve\nabla Q,
  \label{Q}
\eeq
where we henceforth denote the integration $\int d^2r$ over the transverse
directions simply by $\int$, and to first order in $\ve\epsilon$ we replace
the $n_v$ in $Q$ by $n_v^0$.  (The second order contribution from the
dependence of $Q$ on $n_v$ leads to terms of relative order $1/N_v$
~\cite{wgbp}.)  Since $\ve\nabla Q \simeq (\partial Q/\partial n)\ve\nabla n$,
and $\partial Q/\partial n$ is generally positive, vortices are driven towards
regions with lower density.  They are prevented from moving too far, however,
by the rotational kinetic energy of the system which increases as they move
away from a perfect triangular lattice.

    To determine the kinetic energy of flow in terms of $\ve \epsilon$, we
start with the equation for quantization of circulation to relate the
modification of the flow velocity to the vortex displacements:
\beq
    \oint_{{\cal C}(r)}  {\hat{\bf v}}({\bf r'})\cdot d{\ve\ell} =
  \frac{2\pi}{m}N_v(r),
\eeq
where $\hat{\bf v}$ is the microscopic local velocity in the lab frame,
the contour $\cal C$ is a circle of radius $r$ about the origin, and $N_v(r)$
is the number of vortices contained within the circle.  For a uniform lattice
with $N_v(r)$ replaced by $\pi n_v^0 r^2$, the flow is solid body with $v_\phi
= \Omega r$.  Under a local radial displacement of the vortex lattice, $\delta
N_v(r) = -2\pi\epsilon_r rn_v^0$ at fixed $r$.  Thus the smooth flow of the
fluid in the azimuthal direction is changed by
changed by
\beq
   v_{R,\phi} = - \frac{2\pi}{m} n_v^0\epsilon_r,
\label{deltav}
\eeq
which is the fluid flow in the rotating frame.  An outward displacement of
the vortices leads to a slowing down of the rotational flow.  The kinetic
energy of the smoothed flow in the rotating frame is
\bea
  K&=&\frac12\int n(r)m v^2 - \Omega \int n(r)mrv_\phi \nonumber \\
    &=& \frac12\int n(r)m ({\bf v} - r\Omega\hat {\ve \phi})^2 - \frac12 \bar I
    \Omega^2,
\eea
where the integral in the second term of the first line is the angular
momentum of the flow (the energy associated with the variations about smooth
flow is included in $Q$); and $\bar I = \int m n(r)r^2$ is the moment of
inertia calculated for the smoothed density profile.  A distortion of the
vortex lattice at fixed $n(r)$, leads to a second order contribution to the
kinetic energy
\beq
   \delta K = \frac12 \int m n(r) v_{R,\phi}^2,
  \label{K0}
\eeq
which, with (\ref{deltav}), becomes the second order contribution to
$E'$ in terms of $\ve\epsilon$:
\beq
   \delta^{(2)}E' = \int\frac{2\pi^2 }{m}(n_v^0)^2n(r)\epsilon_r^2.
  \label{K2}
\eeq
Corrections to this result are of relative order $1/N_v$.

    The energy to second order also includes contributions arising from
gradients of $\ve\epsilon$, and this has the usual form for an elastic medium,
\bea
    E_{\rm el}(r) = \int\left\{2C_1 ({\ve \nabla}\cdot{\ve\epsilon})^2
  +C_2\left[\left(\frac{\partial \epsilon_x}{\partial x}
 -\frac{\partial\epsilon_y}{\partial y}\right)^2 \right. \right.
  \nonumber\\ \left. \left.
  + \left(\frac{\partial \epsilon_x}{\partial y} +\frac{\partial
  \epsilon_y}{\partial x}\right)^2\right]\right\},
 \label{elastic}
\eea
where $C_1(n)$ is the compressional modulus and $C_2(n)$ the shear modulus
of the vortex lattice \cite{radial}.  Since the $C$'s are of order $\Omega n$,
and $\nabla\epsilon$ is of order $\epsilon/R$, where $R$ is the radius of the
system, the elastic energy density is of order $\Omega n \epsilon^2/R^2$.  By
contrast the kinetic energy density in (\ref{K2}) is of order $m\Omega^2 n
\epsilon^2$ (where $\ell= 1/\sqrt{\pi n_v}$ is the characteristic spacing of
vortices), a factor $\sim R^2/\ell^2 = N_v$ greater than the elastic
energy (\ref{elastic}); thus the elastic energy may be neglected in
determining the distortion of the lattice.

    The two important terms in the change in the energy under a vortex
displacement together are
\beq
    \delta E' = \int n_v^0 \ve\epsilon \cdot \ve\nabla Q(n)+
    \int\frac{2\pi^2 }{m}(n_v^0)^2n(r)\epsilon_r^2.
 \label{deltaE}
\eeq
Minimizing $\delta E'$ with respect to $\epsilon_r$ we find the equilibrium
lattice distortion \cite{anglinread},
\beq
    \epsilon_r = -\frac{1}{4\pi\Omega}\frac{dQ}{dr}
  \label{dQdr}
\eeq
As we now show, $Q\sim n/m$ in the slow and fast rotation limits, so that
$\epsilon \sim - \ell^2 d\ln n/d r$.

\section{Slow rotation}

    In the slowly rotating limit, the kinetic energy of an individual vortex
at position ${\bf R}_j$ is approximately $\int_j {\hat n(\bf r)}/2mr^2 \approx
\gamma n(R_j)$, where $\hat n$ is the microscopic particle density in the
region of the vortex, the integration is over the Wigner-Seitz cell of radius
$\ell$ containing the vortex, $r$ is measured from the center of the vortex,
$\gamma = (\pi/m) \ln(\ell/\xi)$, $\xi$ is proportional to the vortex core
size, and the smoothed density, $n$, is evaluated at the position of the
vortex.  Thus the leading term in the energy of the vortices is
\beq
   E_{\rm vort}= \int n_v(r) n(r) \gamma.
  \label{Egamma}
\eeq
To leading logarithms, $Q = n\gamma$ in this limit,  and
\beq
    \epsilon_r = - \frac{\ell^2}{4}\ln(\ell/\xi)\frac{d\ln n(r)}{dr},
\eeq
in agreement with Refs. \cite{anglin} and \cite{sheehy}.

    Before turning to the rapid rotation limit, we remark that one can
calculate $Q$ in general between the slow rotation and very rapid rotation
limits using the approach of Ref.~\cite{wgbp}.  The calculations, which we do
not do here, are complicated by the presence of explicit non-negligible
contributions to the vortex energy that depend on $dn/dr$ as well as
$d^2n/dr^2$, as well as by the need to include explicitly the local fluid
velocities in the intermediate regime.

\section{Lowest Landau level limit}

    A Bose-condensed system rotating at frequency close to the transverse trap
frequency, $\omega$, can be described by the trial condensate wave function,
\begin{equation}
 \Psi({\bf r})= N^{1/2}h(r) \phi_{\rm LLL},
 \label{ansatz0}
\end{equation}
where $\phi_{\rm LLL}=\chi e^{-r^2/2d^2}$ is composed only of lowest
Landau levels, with $\chi$ a polynomial in $z=x+iy$; $d^2=1/m\omega$, and $N$
is the total number of particles.  As long as the interaction energy per
particle is small compared with $\hbar\omega$, one does not expect the
modification $h$ of the LLL wave function, which admixes higher Landau levels,
to vary significantly over the scale of the intervortex separation.  It is
thus a reasonable approximation to assume that $h$ is a real slowly varying
function dependent only on $r$.

    We first derive the angular momentum in the state (\ref{ansatz0}),
\bea
      \langle L\rangle &=& \int
    \Psi^*({\bf r})\left(z\frac{\partial\Psi({\bf r})}{\partial z}
   -z^*\frac{\partial\Psi({\bf r})}{\partial z^*}\right)
  \nonumber\\
    &=& N\int d^2r \chi^*h^2 e^{-r^2/d^2}z\frac{\partial\chi}{\partial z},
\eea
where in the latter form we use the fact that $(z\partial/\partial z-
z^*\partial/\partial z^*)r^2=0$.  Then integrating by parts we have
\bea
      \langle L\rangle &=& -N\int
    \frac{\partial}{\partial z}\left(z \chi^* h^2 e^{-r^2/d^2}\right)\chi
 \nonumber \\
    &=& -N + \omega I - \int \hat n r \frac{d\ln h(r)} {d r},
  \label{LLLL}
\eea
where $\hat n(\bf r) = |\Psi({\bf r})|^2$ and $I = m\int \hat n r^2$ is
the moment of inertia.  [The final term in Eq.~(\ref{LLLL}) corrects the
expression for $\langle L\rangle$ in Ref.~\cite{wbp}.] More generally, for
real $h({\bf r})=h(x,y)$, the expectation value of the angular momentum is
\bea
  \langle L\rangle &=& \int \hat{n}
 \left\{\left(\frac{r^2}{d^2}-1\right) - {\bf r}\cdot\ve\nabla \ln{h({\bf
  r})}\right\} \nonumber\\
  &=& -N + \omega I - \int \hat{n}\ {\bf r}\cdot\ve\nabla \ln{h({\bf r})}.
\eea

    To calculate the expectation value of the Hamiltonian in the state
(\ref{ansatz0}),
\bea
  E = \int \Psi^*\left(-\frac{\nabla^2}{2m}+\frac12 m \omega^2 r^2
\right)
   \Psi + E_{\rm int},
\eea
where $E_{\rm int}$ is the energy due to interparticle interactions, we
note that for a lowest Landau level,
\beq
   \left(-\frac{1}{2m}\nabla^2+\frac12 m \omega^2 r^2\right)
      \chi e^{-r^2/2d^2}    \nonumber \\
      = \omega \left(z\frac{\partial}{\partial z}-
    z^*\frac{\partial}{\partial z^*}\right)
    \chi e^{-r^2/2d^2}.
\eeq
By a calculation similar to that of $\langle L \rangle$ above, we find
\cite{whoops},
\bea
   E &=& \omega^2 I + \int
   \frac{\hat n}{2m}\left(\frac{d\ln h}{d r}\right)^2
  \nonumber \\
   &&- \omega \int \hat n r \frac{d \ln h(r)}{d r} + E_{\rm int}.
   \label{E}
\eea
For a more general $h({\bf r})$, the energy is
\beq
  E = \omega^2 I + \int \frac{\hat{n}}{2m} [\ve\nabla \ln{h({\bf r})}]^2
  - \omega \int \hat{n}\ {\bf r}\cdot\ve\nabla \ln{h({\bf r})} \nonumber\\
  + E_{\rm int}.
\eeq
Since $d \ln h(r)/dr$ is of order $1/R$, we may replace $\hat n(\bf r)$ by
its coarse grained average $n(r)$ in the terms in $E$ and $\langle L \rangle$
involving $d \ln h(r)/dr$.  The energy $E' = E-\Omega \langle L\rangle$ in the
rotating frame is then
\bea
   E' &=& \omega(\omega-\Omega) I +\Omega N
   +\int
     \frac{n(r)}{2m}\left(\frac{d \ln h}{d r}\right)^2
      \nonumber\\
  && - (\omega-\Omega) \int n(r) r \frac{d \ln h(r)}{d r}
   + \frac{bg}{2}\int n(r)^2,
\nonumber\\
   \label{Eprime}
\eea
where $b$ is the Abrikosov parameter.

    To determine the structure of the condensate it is most convenient to
regard $h(r)$ and the smoothed density $n(r)$ as the independent variables in
the energy (\ref{Eprime}).  Variations of $h(r)$ at fixed $n(r)$ must change
the local vortex density, $n_v(r)$.  To see how, we write the generalization
for the state (\ref{ansatz0}) of the result of Ref.~\cite{jason} relating the
particle density to the vortex positions:
\beq
    \frac14 \nabla^2 \ln\left(\frac{n(r)}{h^2(r)}\right) = -m\omega +
       \pi n_v(r).
 \label{magic}
\eeq
Thus, under a variation of $h(r)$ at fixed $n(r)$,
\beq
    \delta n_v(r) = -\frac{1}{2\pi} \delta \, \nabla^2 \ln h(r).
\eeq
While the moment of inertia, $\bar I$, for the smoothed density is
independent of $h(r)$ at fixed $n(r)$, the difference $I-\bar I$ does depend
on the density of vortices, and, as shown in Ref.~\cite{wgbp}, has the
structure,
\beq
   I-\bar I \sim \int \frac{m}{\pi} \frac{n(r)}{n_v(r)}.
\eeq
Thus under a variation of $h(r)$, we find, after integration by parts, that
\beq
   \delta(I-\bar I\,) \sim \int \frac{1}{m\Omega^2} \frac{dn(r)}{dr}\delta
      \frac{d\ln h}{dr}.
\eeq
This term is of relative order $1/N_v$ compared with the variation of the
term $-(\omega-\Omega) \int n(r) r d \ln h(r)/d r$ in $E'$, and can be
neglected.

    We now minimize the energy (\ref{Eprime}) with respect to $h(r)$ at fixed
smooth density $n(r)$ and find,
\beq
  \ve\nabla \ln h(r) = m(\omega - \Omega){\bf r},
 \label{lnh}
\eeq
so that
\beq
   h(r)=  C e^{m(\omega - \Omega)r^2/2}.
 \label{h}
\eeq
Substituting this result back into Eq.~(\ref{ansatz0}), we obtain
\beq
 \Psi({\bf r}) \sim N^{1/2}\chi e^{-m\Omega r^2/2}.
 \label{ansatz1}
\eeq
Although in terms of the frequency $\omega$, this wave function includes
higher Landau levels via $h$, their only effect is to change the oscillator
frequency in the levels to the rotation frequency, $\Omega$.

    Minimization of $E'$ with respect to $n(r)$, with Eq.~(\ref{h}),
yields the Thomas-Fermi profile,
\bea
     n(r) &=& \frac{1}{bg}\left(\mu - \frac{mr^2}{2}(\omega^2-\Omega^2)
\right) \nonumber \\
          &=& n(0)\left(1-\frac{r^2}{R^2}\right)
\eea
plus terms of relative order $1/N_v$.  Here $\mu$ is the chemical potential. The radius of the cloud is
$R=[8Nbg/\pi m(\omega^2-\Omega^2)]^{1/4}$.  The Thomas-Fermi profile reflects
the reduction of the effective trapping potential by the centrifugal
potential.

    We now derive the distortion of the lattice.  Measuring the displacements
with respect to a uniform triangular lattice of vortex density $n_v^0 =
m\Omega/\pi$, and using use Eq.~(\ref{lnh}) in (\ref{magic}), with
(\ref{deltanv}), we find,
\beq
    \frac14 \nabla^2 \ln \left(\frac{n(r)}{h^2(r)}\right) =
             -m(\omega-\Omega) -n_v^0 \ve\nabla \cdot \ve\epsilon\,(r).
  \label{neps1}
\eeq
For equilibrium lattices, the displacement $\ve\epsilon$ is entirely in the
radial direction, and we may integrate (\ref{neps1}) to find,
\beq
  \frac{d\ln h(r)}{dr} = \frac12 \frac{d\ln n(r)}{dr} +m(\omega-\Omega)r +
    2\pi n_v^0 \epsilon_r.
  \label{neps2}
\eeq
With Eq.~(\ref{lnh}) for $h$, this result reduces to
\beq
    \epsilon_r = -\frac{1}{4m\Omega}\frac{d\ln n}{d r},
   \label{epsLL}
\eeq
which has the form (\ref{dQdr}) with $Q=\pi n/m$.  Using the Thomas-Fermi
profile, we find
\beq
   \epsilon_r = \frac{r}{2m\Omega}\frac{1}{R^2-r^2}.
\eeq

    Were we to assume a uniform lattice ($\ve\epsilon = 0$), then
Eq.~(\ref{neps2}) would imply
\beq
    h(r) \sim \sqrt{n(r)}e^{m(\omega-\Omega)r^2/2},
\eeq
and thus
\beq
    \Psi \sim \sqrt{n(r)}\chi e^{-m\Omega r^2/2}.
\eeq
The admixture of higher Landau levels would not only modify the effective
frequency, but modulate the wave function by the square root of the smoothed
density.  This result for $h(r)$ does not minimize the energy,
due to the imposed constraint of a uniform triangular lattice \cite{omegav}.

\subsection{Elastic energy in the LLL}

    We now compute the terms in the energy associated with the distortion of
the vortex lattice.  Inserting Eq.~(\ref{neps2}) for $d h/d r$ into
Eq.~(\ref{Eprime}), we derive the energy in the rotating frame, written in
terms of the coarse grained density, $n(r)$, and the lattice displacement:
\bea
   E'\{n(r),\epsilon_r(r)\} &=& \frac{\omega^2-\Omega^2}{2} \bar I
      +\omega(\omega-\Omega)(I-\bar I\,)
   + E_{\rm int} \nonumber \\ &&+\Omega\int \epsilon_r\frac{d n}{d r} +
   2m\Omega^2\int  n\epsilon_r^2.
 \label{k1}
\eea
The penultimate term has the same form as that in the slowly rotating
limit, (\ref{Egamma}), only with $\gamma = \pi/m$.  The final term is the same
as in (\ref{deltaE}), and is the modification of the kinetic energy associated
with the decreased azimuthal velocity caused by radial vortex displacements.
Again, minimizing Eq.~(\ref{k1}) with respect to $\epsilon_r$, dropping the
corrections arising from the difference between $I$ and $\bar I$, which are
suppressed by a factor $1-\Omega/\omega$, we are led directly to
(\ref{epsLL}).  The angular momentum in terms of $\epsilon_r$ and $n$ is
\bea
  \langle L \rangle = \omega(I-\bar I\,) +\Omega \bar I -2m\Omega\int
nr\epsilon_r.
\eea

\section{Connection with elastohydrodynamics}

    The results for the equilibrium lattice distortion are all contained in
the elastohydrodynamic theory of
Refs.~\cite{chandler,tkmodes,qhmodes,gifford}, when the dependence of the
elastic energy on density gradients is taken into account.  The key equations
in the elastohydrodynamic theory (all in the rotating frame, denoted by a
subscript $R$) are the conservation of circulation,
\beq
{\bf v}_R + 2{\ve\Omega}\times\ve\epsilon = {\ve \nabla} \Phi/m,
\label{vphase}
\eeq
where $\Phi$ is the superfluid phase; the superfluid acceleration
equation, the time derivative of Eq.\ (\ref{vphase}),
\beq
  m\left({{\partial {\bf v}_R}\over{\partial t}} + 2\ve\Omega \times
    \dot{\ve \epsilon}\right) = - {\ve \nabla} (\mu + V_{\rm eff}),
  \label{supaccel}
\eeq
where $\mu$ is the chemical potential of the matter; and the equation for
conservation of momentum,
\beq
  m\frac{\partial {\bf j}_R}{\partial t} + 2m\ve\Omega\times {\bf j}_R
  +{\ve \nabla} P + n{\ve\nabla} V_{\rm eff} = -\ve \sigma.
  \label{momcons}
\eeq
Here ${\bf v}_R$ and ${\bf j}_R$ are the smoothed superfluid velocity and
current density in the rotating frame, $P$ is the pressure, and $V_{\rm
eff}(r) = V_{\rm trap}(r)-\frac12 m\Omega^2 r^2$.  The elastic force,
$-\ve\sigma$, is given by
\beq
 \ve \sigma({\bf r},t) =  \left.\frac{\delta E'}{\delta \ve
   \epsilon}\right|_{n,{\bf v}_R},
 \label{sigi}
\eeq
where $E'$ is the full energy in the rotating frame for the distorted
lattice, including the usual elastic energy (\ref{elastic}).  In the
elastohydrodynamics, which also describes non-equilibrium dynamics, the
smoothed density, $n({\bf r})$, the displacement, $\ve \epsilon({\bf r})\,$,
and the smoothed superfluid velocity, ${\bf v}_R({\bf r})$, are independent
dynamical degrees of freedom, and one must regard $E'$ as a functional of
these three independent variables.  Only in the determination of $\epsilon$ in
equilibrium can one eliminate the superfluid velocity as an independent
variable.  Keeping ${\bf v}_R$ fixed means that the kinetic energy variation,
$\delta K$, and hence $\delta^{(2)}E'$, Eqs.~(\ref{K0}) and (\ref{K2}), do not
contribute to $\ve\sigma$.

    In equilibrium, where one has axial symmetry, Eq.~(\ref{vphase}) reduces
simply to (\ref{deltav}); then from Eq.~(\ref{supaccel}), $d(\mu+V_{\rm eff})
= 0$.  In addition, at zero temperature, ${\ve \nabla} P = n{\ve \nabla} \mu$.

    To show how the displacement of the equilibrium lattice, (\ref{epsLL}),
emerges from the elastohydrodynamics, we write ${\bf j}_R = n{\bf v}_R$ and
subtract Eq.~(\ref{supaccel}) from (\ref{momcons}) divided by $n$, to find
\cite{encons},
\beq
  2m\ve\Omega\times ({\bf v}_R -\dot {\ve \epsilon}) =
  -\frac{\ve \sigma}{n}.
 \label{veps}
\eeq
The radial component of this equation in equilibrium reads
\beq
  2m\Omega v_{R,\phi} =  \frac{\sigma_r}{n},
  \label{veps1}
\eeq
so that from Eq.~(\ref{deltav}),
\beq
  4m\Omega^2 \epsilon_r =  -\frac{\sigma_r}{n}.
  \label{veps2}
\eeq
Evaluating $\sigma$ from the elastic energy contribution $\Omega\int
\epsilon_r dn/d r$ in Eq.~(\ref{k1}) we are led immediately to the vortex
displacement (\ref{epsLL}) in the LLL.

\acknowledgements

    Author GB is grateful for the hospitality of the Aspen Center for Physics,
where part of this work was carried out.  Author GW was supported in part by
Grants-in-Aid for Scientific Research provided by the Ministry of Education,
Culture, Sports, Science, and Technology through Research Grant No.~14-7939,
by the Nishina Memorial Foundation, and by a JSPS Postdoctoral Fellowship for
Research Abroad.  This research was also supported in part by NSF Grants
PHY03-55014 and PHY05-00914.

\end{document}